# Edge Channels of Broken-Symmetry Quantum Hall States in Graphene probed by Atomic Force Microscopy

Sungmin Kim[1,2*], Johannes Schwenk[1,2*], Daniel Walkup[1,2*], Yihang Zeng[3*], Fereshte Ghahari[1,2], Son T. Le[1,4], Marlou R. Slot[1,5], Julian Berwanger[6], Steven R. Blankenship[1], Kenji Watanabe[7], Takashi Taniguchi[8], Franz J. Giessibl[6], Nikolai B. Zhitenev[1], Cory R. Dean[3†], and Joseph A. Stroscio[1†]

[1]Physical Measurement Laboratory, National Institute of Standards and Technology, Gaithersburg, MD 20899, USA.
[2]Institute for Research in Electronics and Applied Physics, University of Maryland, College Park, MD 20742, USA.
[3]Department of Physics, Columbia University, New York, NY 10027, USA.
[4]Theiss Research, La Jolla, CA 92037, USA.
[5]Department of Physics, Georgetown University, Washington, DC 20007, USA.
[6]Institute of Experimental and Applied Physics, University of Regensburg, Regensburg 93040, Germany.
[7]Research Center for Functional Materials, National Institute for Materials Science, Tsukuba, Ibaraki 305-0044, Japan.
[8]International Center for Materials Nanoarchitectonics, National Institute for Materials Science, Tsukuba, Ibaraki 305-0044, Japan.

**One Sentence Summary:** The elusive topologically protected edge channels of the broken-symmetry quantum Hall states of graphene are spatially and energetically mapped, resolving the real-space energy dispersion of the ground state.

**Abstract:** The quantum Hall (QH) effect, a topologically non-trivial quantum phase, expanded and brought into focus the concept of topological order in physics. The topologically protected quantum Hall edge states are of crucial importance to the QH effect but have been measured with limited success. The QH edge states in graphene take on an even richer role as graphene is distinguished by its four-fold degenerate zero energy Landau level (zLL), where the symmetry is broken by electron interactions on top of lattice-scale potentials but has eluded spatial measurements. In this report, we map the quantum Hall broken-symmetry edge states comprising the graphene zLL at integer filling factors of $\nu = 0, \pm 1$ across the quantum Hall edge boundary using atomic force microscopy (AFM). Measurements of the chemical potential resolve the energies of the four-fold degenerate zLL as a function of magnetic field and show the interplay of the moiré superlattice potential of the graphene/boron nitride system and spin/valley symmetry-breaking effects in large magnetic fields.

The integer QH effect results when a 2-dimensional (2D) electron system is subjected to a perpendicular magnetic field (*1–3*). The metrological precision of the Hall conductance is understood in terms of the topological invariant of the Chern number associated with the Berry connection (*4–7*). The precise quantization of the Hall conductance is related to the absence of backscattering in topologically protected chiral one-dimensional edge states with different momentum directions at the device boundaries. Imaging of QH edge states has been challenging due to their limited spatial extent and their location at the boundaries of the quantum Hall system. A number of notable attempts include: scanning gate microscopy (*8*), scanning single electron transistor (SET) measurements (*9*), scanning force microscopy (*10*, *11*), scanning charge accumulation (*12*), and scanning microwave impedance microscopy (*13*). More recent intriguing progress in imaging quantum Hall edge states has been made using SQUID-on-tip measurements of graphene (*14*), however the authors were not successful in imaging any broken-symmetry states inside the graphene zLL.

---

* These authors contributed equally to this work.
†Corresponding authors: cdean@phys.columbia.edu, joseph.stroscio@nist.gov

The graphene Landau level structure is determined by a combination of the Dirac-like linear energy-momentum dispersion and the $\pi$-Berry phase associated with the Dirac point resulting in Landau energies $\varepsilon_N = \pm\sqrt{2e\hbar v_F^2 B|N|}$, where $e$ is the elementary charge, $\hbar$ is Planck's constant divided by $2\pi$, $v_F$ is the Fermi velocity, $B$ is the magnetic field and $N = 0, \pm 1, ...$ is the Landau level index, resulting in a non-uniform Landau level spacing (*15*, *16*). The zLL in graphene, with orbital index $N = 0$, comprises a set of fourfold degenerate Landau levels that are fixed at the Dirac point in the absence of SU(4) symmetry-breaking effects. The richness of physics in this regime is dominated by the interplay between the Zeeman energy, $E_Z = g\mu_B B$, where $g$ is the electron g-factor and $\mu_B$ is the Bohr magneton, against the sublattice anisotropy of Coulomb interactions which scale as $E_V = \frac{1}{4\pi\epsilon_0\epsilon_r}\frac{ae^2}{l_B^2}$, where $a$ is the lattice constant, $\epsilon_0$ is the vacuum permittivity, $\epsilon_r$ is the relative permittivity, and $l_B^2 = \hbar/eB$ is the magnetic length (*17–23*). The internal degrees of freedom in graphene give rise to a rich variety of ground states for the $N = 0$ Landau level, depending on competing interactions that break the approximate SU(4) isospin symmetry (*17*). Broken isospin symmetry encompassing a variety of different spin and valley orders is observed as additional gapped states at integer filling factors $\nu = 0, \pm 1$ (Fig. 1D). Near the sample boundary, the broken isospin states disperse into positively dispersing (electron-like) and negatively dispersing (hole-like) states, leading to gapped or gapless edge modes depending on specific ground-state symmetries (Fig. 2C) (*18*, *19*, *24*). Imaging the spatial properties of these broken-symmetry states can thus shed light on the competing interactions and makes direct contact with theoretical models.

To image the QH edge states for the present study, we chose to use a dual-gated graphene quantum Hall device where the Hall bar boundary is defined by a lateral pn junction set by two independent tunable back gates (*25–27*). The advantage of this approach is that the boundary is free of major defects with the graphene lattice perfectly continuous across quantum Hall edge interface, as opposed to a physical boundary where the graphene sheet is shaped using harsh treatments such as reactive ion etching, producing many defects along the physical edge. Moreover, the area outside the boundary can be tuned to an "electronic" insulator by setting the graphene density to zero (filling factor $\nu = 0$) where graphene shows an insulating state at high magnetic fields (*28*). Such definition of boundaries have led to superior quantum Hall signatures in various graphene device geometries (*29–31*).

Measurements of macroscopic and microscopic quantum Hall properties were made using a unique instrument that is capable of simultaneous magnetotransport, scanning tunneling microscopy, and AFM measurements on a given device at ultra-low temperatures (*32–34*). The instrument was operated at 10 mK for all measurements with a perpendicular magnetic field up to 15 T. Figure 1A shows an optical image of the dual-gated device [see fig. S1 and (34) for more device details]. The dotted red trapezoid shows the area gated by the local graphite gate G1 defining the interior carrier density. Outside the red trapezoid a second graphite gate G2 defines the exterior density, which is set to zero density to define the Hall bar geometry (fig. S1A). Magnetotransport measurements of the Hall and longitudinal resistances are shown in Figs. 1, B and C. Broken-symmetry states inside the zLL are seen at filling factors $\nu = 0, \pm 1$, marked by quantized plateaus in the Hall resistance ($R_{XY}$). The same broken-symmetry states are observed in microscopic AFM measurements in both the frequency shift (tip-sample force gradient) vs. G1 (Fig. 1E) and sensor oscillation amplitude (energy dissipation) vs. G1 (Fig. 1F) measurements at $B = 15$ T. In both the force and dissipation measurements, the symmetry-breaking states show up as a small increase in frequency shift or decrease in amplitude over a narrow density range around integer filling factors, $\nu = 0, \pm 1, \pm 2, \pm 3$, and $\pm 4$ (Figs. 1, E and F). The AFM response to the symmetry-breaking states derives from the gapped nature of these states and their associated electronic incompressibility. The formation of an incompressible area under the tip apex leads to changes in the system capacitance and resistance, which alters the electrostatic force gradient between tip and sample. This varies the sensor resonance frequency, and hence the detection of the broken-symmetry states (Fig. 1E) [see supplemental text for further discussion (34)]. The reduced electrostatic field between tip and sample leads to a weaker contrast of the broken-symmetry states around a compensated tip-sample bias potential where the sample bias balances the work function difference between the probe tip and sample, which is shown by the solid white line. The Landau levels are observed as plateaus with jumps at integer filling factors. The behavior will be investigated in more detail in Kelvin probe force microscopy (KPFM) measurements, shown below.

Images of the edge states can be obtained by utilizing the AFM signals (Fig. 1, E and F) in spatial measurements across the quantum Hall edge boundary. The edge states are expected to form a series of incompressible and compressible strips (Fig. 2A) near the boundary edge. The strips originate from the Landau levels that are bent by the potential rise at the boundary and are pinned at the Fermi level in a "wedding cake-like" structure due to interaction effects (Fig. 2B) (35, 36). A compressible strip is formed when a partially filled Landau level is at the Fermi level, which is separated by incompressible strips during Landau level transitions. The spatial dispersion (i.e. the change in real-space position of the state as a function of energy) of the edge states depends on spin/valley symmetry configurations, as indicated schematically for the so-called canted antiferromagnetic (CAF), antiferromagnetic (AF), and ferromagnetic (FM) states in Fig. 2C (18, 19, 24).

Figure 2D shows a series of spatial AFM scans of 400 nm by 400 nm over the electrostatically defined boundary of the device. The images resolve the $\nu = \pm 1$ edge states at $B = 10$ T in the AFM dissipation signal at various local-gate potentials across the quantum Hall boundary, indicated in the black circular region in Fig. 1A. The dissipation signal is the excitation voltage needed to keep a constant oscillation amplitude of 2 nm of the AFM qPlus sensor (34). The region outside the edge in the lower region of each image is set to zero density at $\nu = 0$, corresponding to an electronic insulator. The first row of images shows the incompressible strip corresponding to $\nu = -1$ appearing in the local gate voltage range of $-280$ mV $<$ G1 $<$ $-100$ mV. The $\nu = -1$ strip is detected near the boundary at lower potentials and moves up in the images away from the boundary with increasing gate potential (density) as the central region of the Hall bar over the local gate approaches $\nu = -1$, as indicated by the white arrow. The second row in Fig. 2D shows the emergence of the $\nu = 0$ state occurring inside the boundary and merging with the outside $\nu = 0$ region at G1 = 110 mV. Finally, the third row of images shows the electron counterpart of the top row with the $\nu = +1$ incompressible state, first appearing from the top of the image and then drifting toward the boundary edge with increasing gate potential (310 mV < G1 < 430 mV).

The dispersion of the $\nu = \pm 1$ states observed in Fig. 2D points to a configuration with a gapped valley or sublattice-symmetry breaking at $\nu = 0$, which rules out a gapless ferromagnetic ground state in favor of a gapped canted antiferromagnetic or pure antiferromagnetic ground state (*18*, *19*, *24*), as sketched in Fig. 2C. The correspondence to a gapped ground state can be seen by examining the lower energy state for the AF configuration in Fig. 2C, where the spatial movement of the state approaches the boundary with decreasing energy. Imaging this state in the *XY* plane would thus show the state approaching the boundary with decreasing gate voltage (see white arrow) as observed in the top row of Fig. 2D for the $\nu = -1$ incompressible edge strip. Similarly, as the state above the gap in the AF configuration approaches the boundary in the *y*-position, the state is moving up in energy, and hence corresponds to the $\nu = +1$ incompressible strip movement in the lower panel in Fig. 2D. We note that the present data cannot distinguish between a CAF ground state with small canting and the AF state, although the AF state is expected for perpendicular fields due to the larger valley isospin anisotropy compared to the Zeeman energy (*19*, *24*, *37*).

The frequency shift of the AFM qPlus probe, proportional to the force gradient, shows an inverted parabolic profile as a function of applied electrostatic potential, characteristic of the electrostatic forces, as discussed in Fig. 1E [see (*34*) and fig. S4D]. The vertex of the parabolic response occurs when the applied potential compensates the contact potential difference (CPD) between the probe and graphene and allows for measurements of the local chemical potential by KPFM (*38*, *39*). Figure 3A displays KPFM measurements of the CPD as a function of the sample bias vs. back-gate potential G2 for different magnetic fields between 9 T and 15 T. These measurements were made outside the local gated area with the density of both areas kept the same by ramping G1 and G2 together in a fixed ratio (*34*). A series of plateaus and transitions are observed at various back-gate potentials depending on the magnetic field. Each plateau corresponds to the filling of a particular Landau level, whereas the transitions occur at the incompressible states when the Fermi level is being swept in the gaps between the Landau levels. The data in Fig. 3A shows the characteristic graphene Landau level energy structure discussed above, as seen by scaling the sample bias by $\sqrt{B}$ and the gate potential by $B$, as shown in Fig. 3B. The correspondence to the graphene Landau level density of states is indicated by the lineup of the $N = \pm 0,1,2$ Landau levels in Fig. 3C with the plateaus in Fig. 3B.

On close examination of Fig. 3B, the $N = 0$ Landau level plateau shows it contains four distinct smaller plateaus, as shown in the blowup in Fig. 3E for $B = 15$ T. The four plateaus indicate the lifting of the degeneracy of the $N = 0$ Landau level. A large up and down, "N"-shaped excursion in the chemical potential is observed at the integer filling factors separating the plateaus. Note that the excursion is characterized by very sharp jumps in chemical potential over a small change of filling factor as the incompressible state is entered. It is then followed by a linear transition to the lower extrema, and a gradual settling to the next plateau value. The large excursion, on the order of ≈50 meV for $B = 15$ T, is suggestive of interactions playing a strong role. Indeed, in previous SET measurements such large excursions were interpreted due to exchange enhanced openings of the broken-symmetry gaps due to Pauli exclusion (*20*). The sign of the excursion in our measurements, the peak followed by the dip, is opposite to that in previous measurements (*20*), which requires further theoretical investigation.

The direct measurement of the chemical potential by AFM in Fig. 3E yields the energies of the broken-symmetry states at arbitrary filling factor, both when the Fermi level is in the compressible state (plateau) and when it is in the incompressible state (between plateaus), complementing and expanding existing methods. Traditionally, the energies can be determined only at integer filling factors from transport measurements assuming an activated behavior. These energies can vary greatly between different devices, pointing to disorder contributing to the mobility gaps extracted from such activation measurements. The bare symmetry-breaking potential is usually strongly enhanced by exchange and other correlations at filling factors. In contrast to transport measurements, KPFM measures directly the local chemical potential. The energy differences between the chemical potential plateaus shown in Fig. 3F do reflect the strength of the lattice-scale symmetry-breaking potential but the degree of the enhancement is likely smaller than that at integer filling factors. The enhancement is still significant as one sees that the energies are much larger than the Zeeman energy (solid black line) and the largest energy gap across $\nu = 0$ reaches a value of ≈8 meV at 15 T. The energy gaps show distinct low and high field behaviors. Starting with a plateau or even a slight decrease at fields below 8 T, the $\nu = 0$ gap scales with $\sqrt{B}$ at high fields above 8 T. Scaling as $\sqrt{B}$ is consistent with electron interactions playing a dominant role. The $\nu = \pm 1$ energy gaps show much lower energies first increasing at low field and then turning over to approach the Zeeman energy at the highest field of 15 T. Interestingly, the $\nu = 0$ and $\nu = \pm 1$ energy dependencies can be interpreted as an avoided crossing at around 6-8 T, suggesting a possible isospin phase transition (*21*).

The discussions of the many different isospin configurations which are possible in the SU(4) - symmetric $N = 0$ Landau level have been mostly focused on the ground states (*16–18, 25–28*) or excitations (*40–44*) at integer filling factors. The energy gaps in Fig. 3F are measured at filling factors far off integers, but we expect that these compressible states inherit essential ordering from the corresponding states at integer filling factors. The phase diagram proposed for $\nu = 0$ includes configurations involving a ferromagnetic (FM), charge density wave (CDW), canted anti-ferromagnetic (CAF), and Kekulé (KK) state [see Fig. 18 in Ref. (*17*)]. Recent experimental capacitance measurements in tilted magnetic fields have pointed to the CAF state as the ground state for the $N = 0$ Landau level (*19*). For pure perpendicular magnetic fields, the valley isospin anisotropy is much larger than the Zeeman energy leading to the CAF state approaching the AF ground state (*19, 24, 37*), which is consistent with the spatial measurements of the edge channels in Fig. 2.

An additional sublattice-symmetry breaking is caused by a moiré superlattice that can exist due to the rotational misalignment between the graphene and hBN layer, generating a zero-field gap $\Delta_{AB}$. This introduces an additional competing potential which alters the possible ground-state phase diagram to include a partially sublattice polarized (PSP) state in addition to the CDW and CAF states (*21*). Recent measurements have indicated possible isospin phase transitions between these states as a function of magnetic field (*21*). Atomic resolution STM measurements of the device in Fig. 1A does indeed show a moiré period of ≈4.36 nm corresponding to a misalignment of 3.1° of the graphene lattice relative to the hBN underlayer [fig. S7 (*34*)]. Magnetotransport measurements as a function of misalignment angle have shown that a zero-field gap scales with rotation angle (*45*), and a gap value of 5-10 meV can be expected for a misalignment of 3.1°. This value is consistent with the behavior of $\Delta E(\nu = 0)$ in Fig. 3F if the low field trend is extrapolated to zero field. In addition, the apparent avoided crossing seen in Fig. 3F at fields of 6-8 T are suggestive of a possible isospin transition between a CDW to AF phase at intermediate magnetic fields.

We now turn to the large peak-to-valley excursion in Figs. 3, D and E. While the $\nu = 0$ state is dominated by valley/sublattice anisotropic interactions, the $\nu = \pm 1$ incompressible states are expected to involve mostly spin-polarized states. In all cases, the analysis of energy gaps and line shapes must include the physics of generalized isospin skyrmions (*40–44*). The characteristic energy scale can be as large as $\Delta_0 = \frac{1}{4\pi\epsilon_0\epsilon_r}\sqrt{\frac{\pi}{2}}\frac{e^2}{l_B} \cong 54$ meV, with $\epsilon_r = 5$ at $B = 15$ T. This is consistent with the peak-to-valley excursion in Fig. 3D.

The large excursion seen at integer filling factors in the chemical potential in Fig. 3E is useful as a fingerprint for spatial mapping of the incompressible states in the zeroth Landau level with improved precision and confidence (Fig. 4). Compared to the dissipation measurement in Fig. 2, the essential advantage of the KPFM is that the measurement is compensating the CPD between tip and sample and thus minimizing gating effects in the graphene 2DEG. For the dissipation, as well as the frequency-shift measurements in Figs. 1 and 2 the tip-sample bias is controlled by the experimentalist and not via a KPFM feedback loop, leading to a non-zero electrostatic force between the probe and the sample. Figure 4 shows the KPFM measurements at $B = 10$ T of the chemical potential across the QH edge boundary as a function of *y*-position and local gate potential, G1. At a distance of about 300 nm from the boundary, the different chemical potential plateaus in the top of Fig. 4A correspond to the $N = 0, \pm 1, \pm 2$ Landau levels, as in Fig. 3B. As the probe approaches the QH edge boundary, the electron (hole) states disperse to positive (negative) densities. Inside the $N = 0$ Landau level, the $\nu = \pm 1$ incompressible states are observed to follow a similar dispersion, as seen in the higher-resolution measurement in Fig. 4B [see fig. S5 for additional data (*34*)]. Figure 4C shows the spatial width of the incompressible strips is very narrow on the order of ≈40 nm, consistent with the length scale of the electrostatic potential of the graphite back gates defining the quantum Hall edge [see supplemental text (*34*) for further data and analysis of incompressible strips]. Consistent with the behavior in Fig. 2, the $\nu = -1$ incompressible edge channel is seen dispersing to the left of $\nu = 0$ and $\nu = +1$ disperses to the right. This reaffirms that the dispersion is characteristic of an AF or CAF with small canting ground state with a gapped edge state (*18, 19, 24*).

In summary, we have shown the spatial and energy mapping of the zLL graphene QH states using atomic force microscopy. Spatial mapping of the broken-symmetry edge states was shown using both the AFM dissipation signal and chemical potential measurements across the QH edge boundary. The chemical potential measurements allowed the Landau level energies to be measured both when the Fermi level is in the Landau levels (compressible) and at the integer filling factors (incompressible), not attainable by transport measurements. The advantage of combining macroscopic and microscopic measurements of this graphene system shows promise in future studies of quantum materials, where a multi-mode "Swiss army knife" type instrument is expected to cut through and reveal the underlying physics to harness for future applications.

**Acknowledgments:** We thank A. MacDonald, A. Young, and B. Feldman for useful discussions. We also thank David Goldhaber-Gordon and Derrick Boone for assistance in device fabrication, and William Cullen for technical assistance. **Funding:** J.S., S.K., D.W., and F.G., acknowledge support under the Cooperative Research Agreement between the University of Maryland and the National Institute of Standards and Technology (NIST), Grant No. 70NANB14H209, through the University of Maryland. M.R.S. acknowledges support under the Cooperative Research Agreement between the Georgetown University and NIST, Grant No. 70NANB18H161, through the NIST/Georgetown PREP program. S.T.L. acknowledges support by NIST and grant 70NANB16H170. J.B and F.J.G. acknowledges support by Deutsche Forschungsgemeinschaft, SFB1277, project A02. K.W. and T.T. acknowledge support from the Elemental Strategy Initiative conducted by the MEXT, Japan, Grant Number JPMXP0112101001, JSPS KAKENHI Grant Numbers JP20H00354 and the CREST(JPMJCR15F3), JST. C.R.D. acknowledges support under the Army Research Office Grant No. W911NF-17-1-0323. **Author contributions:** S.K., J.S., D.W., F.G., M.R.S., J.A.S, N.Z. performed the experiments. Y.Z., F.G., S.T.L. designed and fabricated the graphene device. J.B., S.B., F.J.G., and J.A.S. constructed parts of the instrumentation. K.W. and T.T. grew the hBN crystals used in the graphene device. All authors contributed to writing the manuscript. **Competing interests:** F.J.G. holds patents on the qPlus sensor.  **Data and materials availability:** All data is available in the main text or the supplementary materials.

Supplementary Materials:

Materials and Methods

Supplementary Text

Figures S1-S6

References (*46,47*)

**Figure Captions:**

**Fig. 1 Correlation of graphene broken isospin states in macroscopic vs. microscopic measurements.** (**A**) Optical micrograph of the graphene quantum Hall device. The Hall bar edges are defined by a local graphite back gate, G1, underlying the area outlined in the red dashed line, and a global graphite back gate, G2, under the entire Hall bar device (see Ref. (*34*) for further details). The boundary between G1 and G2 defines the quantum Hall edge boundary along the red dashed line. The black circle shows the location for spatial maps across the boundary shown in Fig. 2 and Fig. 4. Magneto-transport measurements of (**B**) the Hall resistance, $R_{XY}(B)$, and (**C**) the longitudinal resistance $R_{XX}(B)$. Filling factors $\nu$ are indicated in white numerals. In both measurements broken-symmetry states in the zeroth Landau level are observed at $\nu = \pm 1$. (**D**) Schematic of the graphene Landau level density of states indicating the four-fold degeneracy due to valley and spin inside each main Landau level. Microscopic atomic force spectroscopy measurements revealing the broken-symmetry states in, (**E**) AFM frequency shift measurements, and (**F**) simultaneously obtained oscillation amplitude signal with constant excitation of 520 mV as a function of sample bias and local gate at $B = 15$ T. A smooth background was subtracted from the data in (E) to enhance the contrast of the broken-symmetry states [see fig. S3 and supplemental text (*34*)]. The white line indicates the zero-contact potential difference (i.e. chemical potential) obtained from a parabolic fit to the frequency shift data vs. sample bias [see fig. S4 (*34*)]. The white numerals indicate the filling factor. All measurements were made at $T = 10$ mK.

**Fig. 2 Spatial mapping of graphene broken-symmetry quantum Hall edge states using AFM dissipation measurements.** (**A**) Schematic of bulk (closed cyclotron orbits with cyclotron frequency $\omega_C$) and edge quantum Hall states leading to compressible and incompressible strips at the device edge boundary. (**B**) Schematic of the "wedding cake-like" series of plateaus in Landau levels near a boundary edge. A compressible strip is formed when a LL is at the Fermi level, separated by incompressible strips during Landau level transitions. (**C**) Schematic of the spatial energy dispersion for three isospin components leading to various ground-state configurations, including a canted antiferromagnetic state (CAF), antiferromagnetic state (AF), and a ferromagnetic state (FM) (*18*, *19*, *24*). The spin direction of the bands projected onto the magnetic field direction is indicated by color: red, aligned; blue, anti-aligned; black, zero net spin along the field direction. (**D**) Series of spatial AFM excitation channel maps (for constant sensor oscillation amplitude of 2 nm) across the quantum Hall edge boundary (indicated in the black circle in Fig. 1A) as a function of the local back gate potential G1. The gate potential outside the boundary was set to $\nu = 0$, an electronic insulator. (first row) The incompressible $\nu = -1$ edge state appears out of the boundary and drifts up with increasing gate voltage. (second row) The $\nu = 0$ appears at the local gate voltage of G1 = 110 mV. (third row) The

incompressible $\nu = +1$ edge state appears from the top of the frame and drifts down with increasing gate voltage. AFM settings: $V_B = -600$ mV, $\Delta f = -2$ Hz.

**Fig. 3 Resolving the energies of the four isospin components of the graphene zero Landau level with Kelvin probe spectroscopy**. (**A**) Kelvin probe measurements varying the sample bias and simultaneously gates G1 and G2 for measurements made outside the Hall bar area. A staircase of plateaus shows various Landau levels occurring at different chemical potentials for various magnetic fields. (**B**) Chemical potential vs. filling factor given by the data in (A) collapsed onto a universal curve by scaling the sample bias by the graphene Landau level energy field dependence along the vertical axis, $E_N \propto \sqrt{B}$, and by the $B^{-1}$ along the horizontal axis to give a density/filling factor axis. Each Landau level is observed by a plateau in the scaled chemical potential. Notice the zero Landau level at zero chemical potential consists of four separate small plateaus indicating the lifting of the four-fold degeneracy. (**C**) The Landau level density of states calculated using the expression in the main text with $B = 1$ T and $v_F = 1.13 \times 10^6$ m/s to fit the locations of the plateaus in (B). (**D**) Blow up of the large up and down excursion in chemical potential at $\nu = -1$ and $B = 15$ T from (E). (**E**) Blow up of the chemical potential of the zeroth Landau level at $B = 15$ T from (B) showing four individual chemical potential plateaus separated by large up/down excursions at the incompressible filling factors, $\nu = 0, \pm 1$. A blow up of the excursion for $\nu = -1$ is shown in (D). The red dashed lines indicate the calculated difference in chemical potentials $\Delta E(\nu = 0, \pm 1)$. (**F**) Calculated energy differences from the chemical potential plateaus in (E) for the $\nu = 0$ (red circles) and $\nu = -1$ / $\nu = +1$ (orange triangles/green squares) filling factors. The values are averaged chemical potential difference values from $\nu - 0.75$ to $\nu - 0.25$ of each integer $\nu$, and the error bars are corresponding one standard deviation. The solid black line shows the Zeeman energy, $g\mu_B B$, with $g = 2$. The solid red line is a fit for $\nu = 0$ data values to $\sqrt{B}$ for values $\geq 8$ T, and the blue line is a linear fit for $B$ values $\leq 8$ T. AFM settings: 5.8 nm oscillation amplitude, $\Delta f = -450$ mHz, 5 Hz bias modulation, except a 1 Hz bias modulation was used for 4 T and 5 T data.

**Fig. 4 Spatial dispersion of graphene broken-symmetry edge states at the quantum Hall edge boundary**. (**A**) Kelvin probe maps at $B = 10$ T of the chemical potential as a function of y-position across the quantum Hall boundary and local gate potential. In the local gated area Landau levels from $N = -2$ [LL(−2)], to $N = +2$, [LL(+2)], are seen in the different colored plateaus. Incompressible signatures due to the large excursions in the chemical potential (see Fig. 3D) are observed inside the $N = 0$ Landau level corresponding to filling factors $\nu = 0, \pm 1$. (**B**) Higher resolution Kelvin probe map of the $N = 0$ Landau level showing that at the quantum Hall edge boundary the $\nu = \pm 1$ channels disperse away from the $\nu = 0$ center line consistent with a gapped AF or CAF with small canting ground state depicted in Fig. 2C. AFM settings: 2 nm oscillation amplitude, $\Delta f = -2$ Hz, 20 mV bias modulation. (**C**) Line trace along **a-a'** in (A) at a local gate potential of G1 $= -1.65$ V. The excursions in the chemical potential indicate incompressible strips of ≈35 nm width at filling factors $\nu = -2$ and $\nu = -1$. The transition at $\nu = -2$ separates the plateaus of the LL(−1) and LL(0) Landau levels. [see fig. S5 for additional data (*34*)].

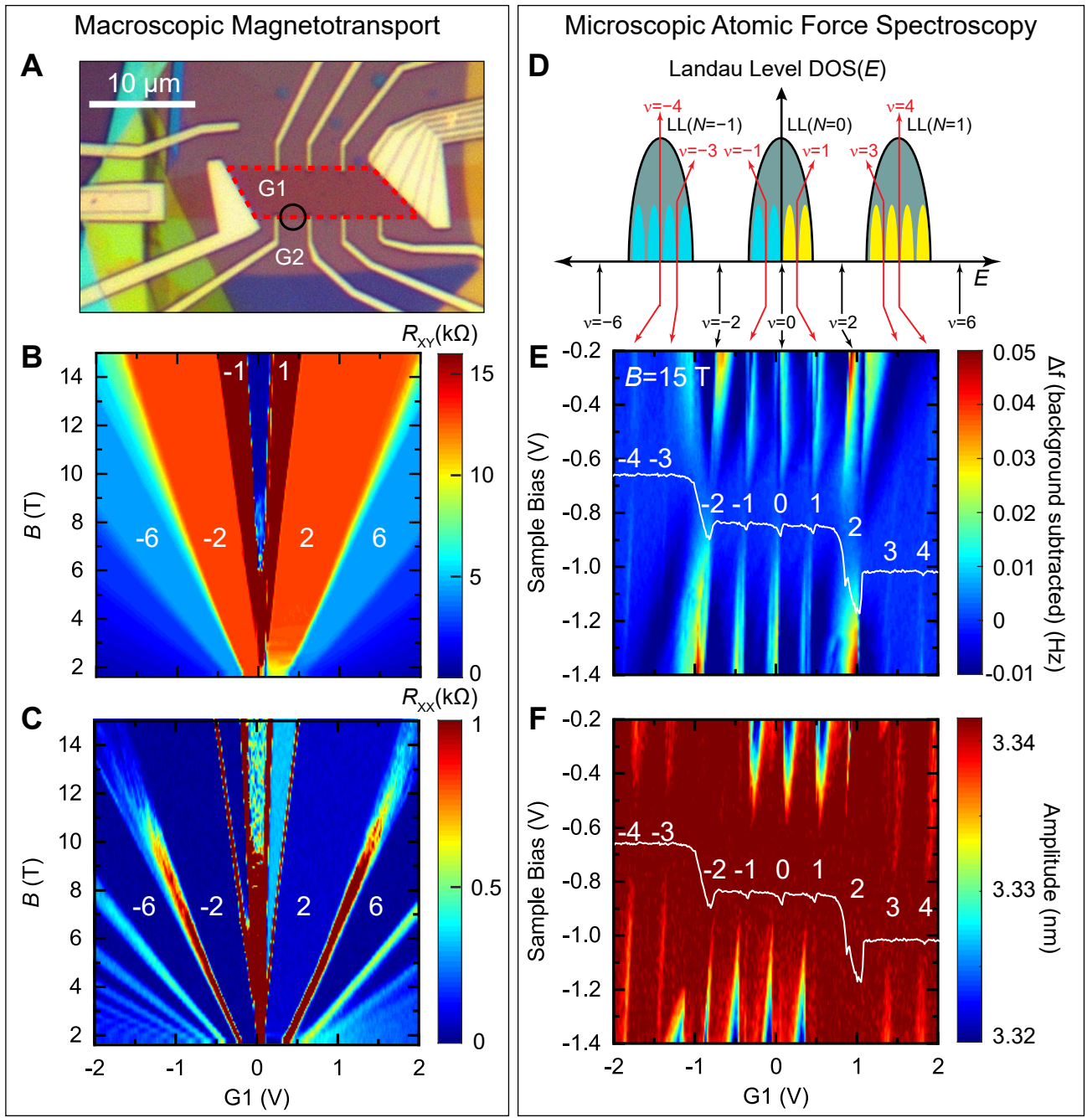

**Fig. 1 Correlation of graphene broken isospin states in macroscopic vs. microscopic measurements.** (A) Optical micrograph of the graphene quantum Hall device. The Hall bar edges are defined by a local graphite back gate, G1, underlying the area outlined in the red dashed line, and a global graphite back gate, G2, under the entire Hall bar device (see Ref. (*34*) for further details). The boundary between G1 and G2 defines the quantum Hall edge boundary along the red dashed line. The black circle shows the location for spatial maps across the boundary shown in Fig. 2 and Fig. 4. Magneto-transport measurements of (B) the Hall resistance, $R_{XY}(B)$, and (C) the longitudinal resistance $R_{XX}(B)$. Filling factors $\nu$ are indicated in white numerals. In both measurements broken-symmetry states in the zeroth Landau level are observed at $\nu = \pm 1$. (D) Schematic of the graphene Landau level density of states indicating the four-fold degeneracy due to valley and spin inside each main Landau level. Microscopic atomic force spectroscopy measurements revealing the broken-symmetry states in, (E) AFM frequency shift measurements, and (F) simultaneously obtained oscillation amplitude signal with constant excitation of 520 mV as a function of sample bias and local gate at $B = 15$ T. A smooth background was subtracted from the data in (E) to enhance the contrast of the broken-symmetry states [see fig. S3 and supplemental text (*34*)]. The white line indicates the zero-contact potential difference (i.e. chemical potential) obtained from a parabolic fit to the frequency shift data vs. sample bias [see fig. S4 (*34*)]. The white numerals indicate the filling factor. All measurements were made at $T = 10$ mK.

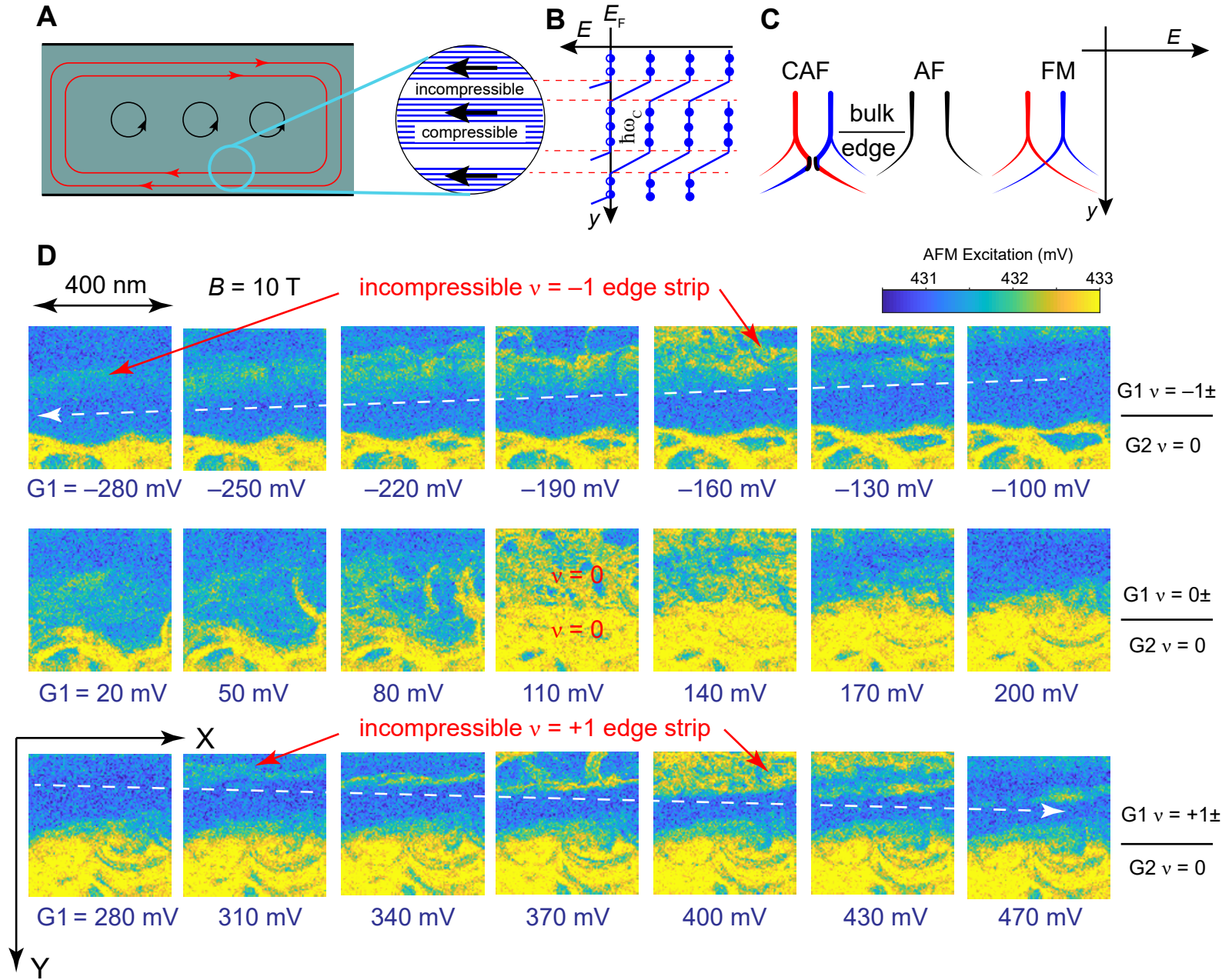

**Fig. 2 Spatial mapping of graphene broken-symmetry quantum Hall edge states using AFM dissipation measurements**. (**A**) Schematic of bulk (closed cyclotron orbits with cyclotron frequency $\omega_c$) and edge quantum Hall states leading to compressible and incompressible strips at the device edge boundary. (**B**) Schematic of the "wedding cake-like" series of plateaus in Landau levels near a boundary edge. A compressible strip is formed when a LL is at the Fermi level, separated by incompressible strips during Landau level transitions. (**C**) Schematic of the spatial energy dispersion for three isospin components leading to various ground-state configurations, including a canted antiferromagnetic state (CAF), antiferromagnetic state (AF), and a ferromagnetic state (FM) (*18, 19, 24*). The spin direction of the bands projected onto the magnetic field direction is indicated by color: red, aligned; blue, anti-aligned; black, zero net spin along the field direction. (**D**) Series of spatial AFM excitation channel maps (for constant sensor oscillation amplitude of 2 nm) across the quantum Hall edge boundary (indicated in the black circle in Fig. 1A) as a function of the local back gate potential G1. The gate potential outside the boundary was set to $\nu = 0$, an electronic insulator. (first row) The incompressible $\nu = -1$ edge state appears out of the boundary and drifts up with increasing gate voltage. (second row) The $\nu = 0$ appears at the local gate voltage of G1 = 110 mV. (third row) The incompressible $\nu = +1$ edge state appears from the top of the frame and drifts down with increasing gate voltage. AFM settings: $V_B = -600$ mV, $\Delta f = -2$ Hz.

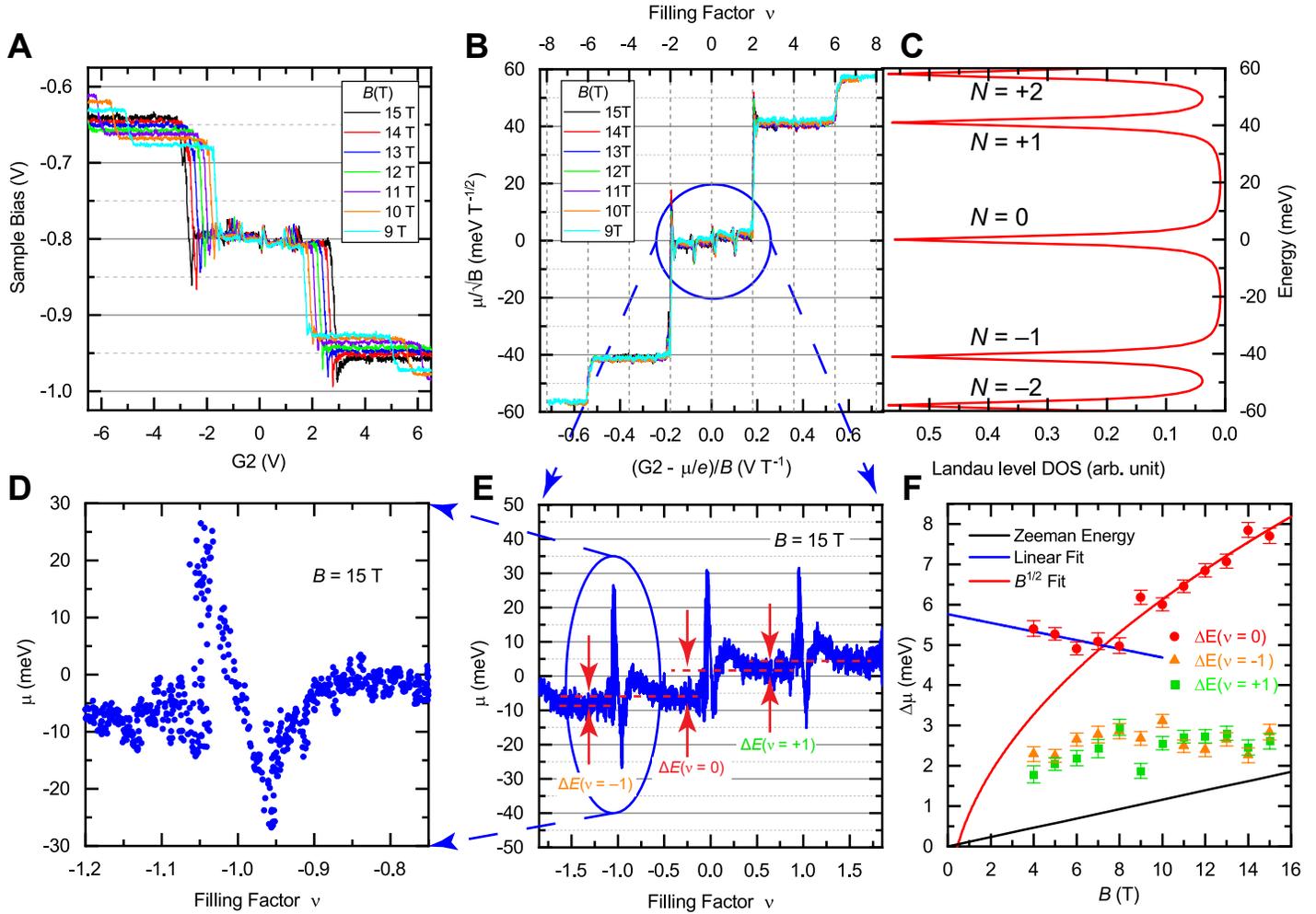

**Fig. 3 Resolving the energies of the four isospin components of the graphene zero Landau level with Kelvin probe spectroscopy**. (**A**) Kelvin probe measurements varying the sample bias and simultaneously gates G1 and G2 for measurements made outside the Hall bar area. A staircase of plateaus shows various Landau levels occurring at different chemical potentials for various magnetic fields. (**B**) Chemical potential vs. filling factor given by the data in (A) collapsed onto a universal curve by scaling the sample bias by the graphene Landau level energy field dependence along the vertical axis, $E_N \propto \sqrt{B}$, and by the $B^{-1}$ along the horizontal axis to give a density/filling factor axis. Each Landau level is observed by a plateau in the scaled chemical potential. Notice the zero Landau level at zero chemical potential consists of four separate small plateaus indicating the lifting of the four-fold degeneracy. (**C**) The Landau level density of states calculated using the expression in the main text with $B = 1$ T and $v_F = 1.13 \times 10^6$ m/s to fit the locations of the plateaus in (B). (**D**) Blow up of the large up and down excursion in chemical potential at $v = -1$ and $B = 15$ T from (E). (**E**) Blow up of the chemical potential of the zeroth Landau level at $B = 15$ T from (B) showing four individual chemical potential plateaus separated by large up/down excursions at the incompressible filling factors, $v = 0, \pm 1$. A blow up of the excursion for $v = -1$ is shown in (D). The red dashed lines indicate the calculated difference in chemical potentials $\Delta E(v = 0, \pm 1)$. (**F**) Calculated energy differences from the chemical potential plateaus in (E) for the $v = 0$ (red circles) and $v = -1 / v = +1$ (orange triangles/green squares) filling factors. The values are averaged chemical potential difference values from $(v - 0.75)$ to $(v - 0.25)$ of each integer $v$, and the error bars are corresponding one standard deviation. The solid black line shows the Zeeman energy, $g\mu_B B$, with $g = 2$. The solid red line is a fit for $v = 0$ data values to $\sqrt{B}$ for values $\geq 8$ T, and the blue line is a linear fit for $B$ values $\leq 8$ T. AFM settings: 5.8 nm oscillation amplitude, $\Delta f = -450$ mHz, 5 Hz bias modulation, except a 1 Hz bias modulation was used for 4 T and 5 T data.

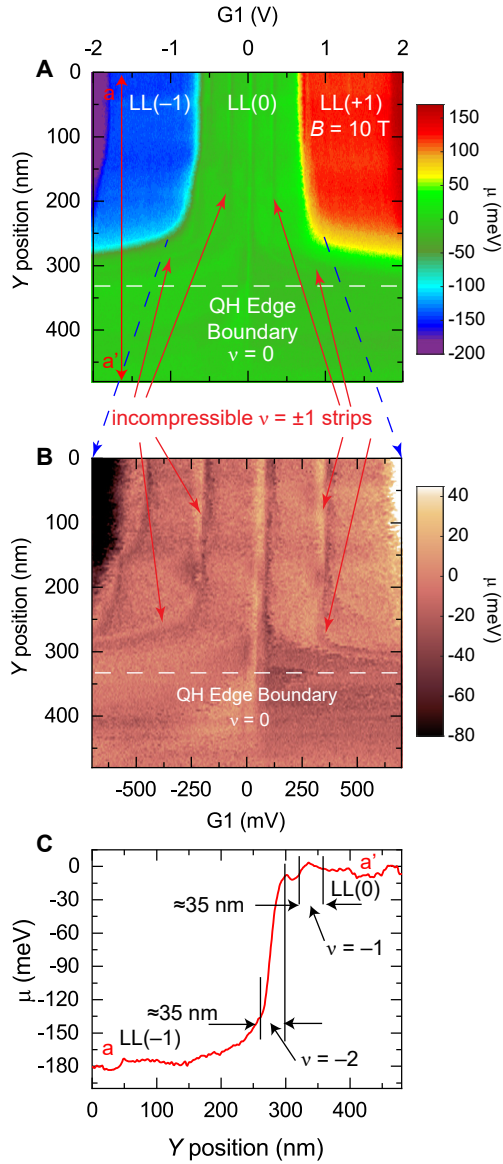

Fig. 4 Spatial dispersion of graphene broken-symmetry edge states at the quantum Hall edge boundary. (A) Kelvin probe maps at $B = 10$ T of the chemical potential as a function of y-position across the quantum Hall boundary and local gate potential. In the local gated area Landau levels from $N = -2$ [LL(−2)], to $N = +2$, [LL(+2)], are seen in the different colored plateaus. Incompressible signatures due to the large excursions in the chemical potential (see Fig. 3D) are observed inside the $N = 0$ Landau level corresponding to filling factors $\nu = 0$, ±1. (B) Higher resolution Kelvin probe map of the $N = 0$ Landau level showing that at the quantum Hall edge boundary the $\nu = \pm 1$ channels disperse away from the $\nu = 0$ center line consistent with a gapped AF or CAF with small canting ground state depicted in Fig. 2C. AFM settings: 2 nm oscillation amplitude, $\Delta f = -2$ Hz, 20 mV bias modulation. (C) Line trace along a-a' in (A) at a local gate potential of G1 = −1.65 V. The excursions in the chemical potential indicate incompressible strips of ≈35 nm width at filling factors $\nu = -2$ and $\nu = -1$. The transition at $\nu = -2$ separates the plateaus of the LL(−1) and LL(0) Landau levels. [see fig. S5 for additional data (34)].

# Supplementary Materials for

# Edge Channels of Broken-Symmetry Quantum Hall States in Graphene probed by Atomic Force Microscopy

Sungmin Kim[1,2*], Johannes Schwenk[1,2*], Daniel Walkup[1,2*], Yihang Zeng[3*], Fereshte Ghahari[1,2], Son T. Le[1,4], Marlou R. Slot[1,5], Julian Berwanger[6], Steven R. Blankenship[1], Kenji Watanabe[7], Takashi Taniguchi[8], Franz J. Giessibl[6], Nikolai B. Zhitenev[1], Cory R. Dean[3†], and Joseph A. Stroscio[1†]

[1]Physical Measurement Laboratory, National Institute of Standards and Technology, Gaithersburg, MD 20899, USA.
[2]Institute for Research in Electronics and Applied Physics, University of Maryland, College Park, MD 20742, USA.
[3]Department of Physics, Columbia University, New York, NY 10027, USA.
[4]Theiss Research, La Jolla, CA 92037, USA.
[5]Department of Physics, Georgetown University, Washington, DC 20007, USA.
[6]Institute of Experimental and Applied Physics, University of Regensburg, Regensburg D-93040, Germany.
[7] Research Center for Functional Materials, National Institute for Materials Science, Tsukuba, Ibaraki 305-0044, Japan.
[8]International Center for Materials Nanoarchitectonics, National Institute for Materials Science, Tsukuba, Ibaraki 305-0044, Japan

†Correspondence to: cdean@phys.columbia.edu, joseph.stroscio@nist.gov

**This PDF file includes:**

Materials and Methods
Supplementary Text
Figs. S1 to S6

**Other Supplementary Materials for this manuscript include the following:**

Data S1 (zipped archive)

---

[*] These authors contributed equally to this work.

**Materials and Methods**

Graphene Device Structure and Fabrication

Figure S1 shows a schematic cross-section of the graphene device heterostructure (Fig. S1A) and an optical top view (Fig. S1B). Two single-crystal graphite gate electrodes and single-crystal hBN dielectrics are employed for optimal sample quality. The two graphite backgate regions G1 and G2 are outlined in Fig. S1B. G1 defines the carrier density of the local interior area as indicated by the red dashed line in Fig. S1B, while G2 defines the carrier density in the outer region (blue dashed line). A quantum Hall boundary edge can thus be generated at the edge of the local gate G1.

The heterostructure is assembled from top to bottom (starting from the global graphite gate G2 as the top layer) using the van der Waals transfer technique so that the bottom of the graphene flake as well as the hBN dielectrics remain free from contamination during the stacking and subsequent fabrication processes. It is then flipped upside down to expose the graphene surface and deposited onto a 285nm $SiO_2$/ $Si^{++}$ substrate before vacuum annealing to remove the polymer film underneath the stack. Electrical connections to the graphene sheet and graphite gate electrodes were made by deposition of the Cr/Pd/Au (2/50/50 nm) metal edge-contacts.

All but one electrode contacting the graphene are in contact with both the G1 region and the G2 region. The one outside contact is used to ensure that the G2 region is in the $\nu = 0$ gapped state during electrical transport and scanning probe measurements. A fan-shaped pattern with gold ridges of 90 nm height is connected to the drain electrode to the graphene sheet for navigation purposes (Fig. S1B). After introducing the device sample to the UHV chamber of the STM instrument it is annealed at a temperature of about 690 K for ≈3 hours to obtain the required cleanliness for STM/AFM measurements.

Navigating to the Device with Scanning Probes

Navigating to the central device area with scanning probes is always a difficult challenge. For this purpose, a fan-shaped pattern extending to ≈500 µm at its widest region is utilized. Figure S2 outlines the procedure for navigating to the device area. First, the probe tip is aligned onto the fan-shaped area using an optical telescope while the STM module is in the upper ultra-high vacuum chamber at room temperature (Fig. S2A). Using the probe tip reflection, a tip-sample gap of the order of 100 µm is set at room temperature. The module is then lowered and locked-in into the dilution refrigerator multi-mode SPM system where it is cooled to a temperature of 10 mK (*32,33*). The landing region is scanned, and a ridge is identified after approaching the fan-shaped runway surface. STM tunneling current or AFM frequency shift feedback can be used for the approach; for this device AFM feedback was used for approach and navigation to minimize the degradation of the probe tip due to interacting with the surface. Once a ridge on the fan-shaped area is found, an automated algorithm is used to follow a given ridge to the device area (Fig. S2B). This algorithm alternates stepping along the ridge direction and quickly imaging the ridge in a "W"-shaped scan. After each "W" scan, the *XY* piezo motor parameters are adjusted to keep the walking direction along the ridge. The successful application of this routine is shown in in Fig. S2C, where AFM traces of the "W" line scans are shown. At certain key places, full AFM scans are made to verify marker features in the devices, as shown in Fig. S2C. After successful navigation, the device region is located as verified by AFM scans of the pattern boundary (Fig. S2D). Further navigation is then performed to check the device area and locate an area for edge studies, as indicated by the black circle in Fig. 1A of the main text.

Multi-Mode STM, AFM, and Transport Instrumentation

The study described in this report is the first to use a newly commissioned multi-mode system with the capabilities of simultaneous AFM, STM, and magnetotransport measurement (*33*). The system utilizes a dilution refrigerator which operates at a base temperature of 10 mK with magnetic fields up to 15 T perpendicular to the sample plane (*32*). Multi-mode measurements are accomplished by using custom designed sample and probe tip holders which feature eight electrical contacts for devices and probe sensors. Magnetotransport measurements were performed using a lock-in amplifier at 25 Hz with a 10 nA source current. The qPlus AFM sensor was a new design which incorporated an integrated excitation electrode on the sensor (*33*). The qPlus sensor had a quality factor of $Q = 1.3 \times 10^5$ and a resonance frequency of $f_0 =$ 23.4 kHz at zero magnetic field. For the qPlus sensor, four contacts were wired, two to read out the AFM sensor, one for the STM tunneling current, and one for the sensor excitation. All eight electrical contacts of the device were utilized. Figure S1C shows the contacts used for the magnetotransport measurement; contacts 1 and 2 were used for measuring the longitudinal resistance and contacts 2 and 3 were used for the Hall resistance shown in Fig. 1, B and C of the main text, respectively. For AFM and KPFM measurements, we used the frequency modulation mode with an oscillation amplitude of 2 nm to 5 nm. For KPFM, we used 1 Hz to 5 Hz modulation on the sample bias voltage.

**Supplementary Text**

AFM Frequency Shift Measurement of Broken Symmetry States

AFM measurements of the frequency shift and dissipation were both sensitive to the occurrence of the broken symmetry states, and Landau levels in general, as shown in Fig. 1 of the main text. The sensitivity originates from the frequency shift caused by the unbalanced Coulomb forces between the tip and the sample:

$$\Delta f_C \sim -\frac{d^2 C(f_0)}{dz^2}(V_\text{B} - V_\text{CPD})^2 \qquad S1$$

where $C(f_0)$ is the capacitance between the tip and the sample at the sensor resonance frequency $f_0$, $V_\text{CPD}$ is the contact potential difference and $V_\text{B}$ is the sample bias. The typical frequency shift curves are shown in Fig. S4D. One can see, for example, that at $V_\text{B} - V_\text{CPD} \cong 0.5$ V along the horizontal axis in Fig. S4D, the frequency shift is $\approx -0.15$ Hz.

The additional positive frequency shift at integer filling factors corresponding to the broken symmetry states seen in Fig. S3C and S3D derives from the openings of gaps which change the resistance and, as a result, the capacitance $C(f_0)$ of the graphene system at the tip location.

An additional dissipation develops once the complex capacitance $C(f_0)$ acquires a phase lag resulting from a large local resistance of the incompressible region. As the resistance grows further, the capacitance $C(f_0)$ decreases, correspondingly causing a positive frequency shift contribution, while the phase lag, and respectively the additional dissipation vanishes. The experimentally observed positive frequency spikes at the integers can be as large as 10-15% of the total frequency shift from the Coulomb attraction (≈0.02 Hz / 0.15 Hz).

In the simplified analysis above, all the changes in $\Delta f_C$ were assigned to the changes in $C(f_0)$ assuming $V_\text{CPD}$ and $V_\text{B}$ being constant. Fig. S3A illustrates that over a larger parameter range, the latter two variables contribute most significantly. The frequency shift at the broken-symmetry gaps is a small signal on top of a large background due to the larger frequency shifts caused by changes in $V_{\text{CPD}}$ originating from the normal cyclotron gaps at filling factors of $\nu = \pm 2$ (along the gate voltage axis), as well as by $V_\text{B}$ (the sample bias axis), both determining the total electrostatic force contributions. We followed Refs. *(46,47)* and subtracted a smoothly-varying background (black curve) from each frequency shift curve (red and blue curves) and plot the residuals, as shown in Figs. S3, B and C. The smooth background averages the original data using a Gaussian filter with a sigma of 0.2 V. Each residual curve is then built into a new frequency shift map, as shown in Fig. S3D and Fig. 1E.

As mentioned above, the frequency shift data contains contributions from electrostatic forces which give rise to a downward parabolic dependence on the applied sample bias (Fig. S4D). The vertex of the parabolic response occurs when the applied potential compensates the contact potential difference (CPD) between the probe and graphene, as illustrated in Figs. S4, A to C. By measuring changes in the CPD we can obtain a measure of the local chemical potential, which responds to changes in Fermi-level position with gate bias, as shown in Fig. 1E of the main text. The measurements of the chemical potential in Fig. 1E were obtained by fitting the parabolic dependence on the sample bias over a window of 300 mV about the vertex. A higher-precision measurement with reduced tip-gating effects is obtained by modulating the sample bias and using lock-in detection to measure the sample bias values compensating the CPD. The force at the modulation frequency ω is:

$$F_\omega = -\frac{1}{2}\frac{dC}{dz}(V_{\text{B+AC}} - V_{\text{CPD}})^2 \sim -\frac{dC}{dz}(V_\text{B} - V_{\text{CPD}})V_{\text{AC}}. \qquad S2$$

Correspondingly, $\Delta f_\omega \sim \frac{dF_\omega}{dz}$ is nullified when $V_\text{B} - V_{\text{CPD}} = 0$.

Incompressible Strips

Figure 4 of the main text showed the dispersion of the broken-symmetry states of the zLL and higher LLs across the QH boundary. Similar data is observed at all magnetic fields investigated as shown in additional data in Fig. S5A for $B = 5$ T. Fig 5B shows two line scans of the chemical potential across the QH boundary at G1 $= -0.9$ V and $-0.6$ V. For the larger gate voltage, we observe three incompressible strips corresponding to filling factor $\nu = -6, -2,$ and $-1$. The width of the $\nu = -6$ and $\nu = -2$ which occurs at the change of the Landau levels are on the order of ≈40 nm, similar or slightly larger than observed for $B = 10$ T in Fig. 4 of the main text. The width of the $\nu = -1$ strip is much narrower on the order of ≈20 nm.

The staircase shape of the chemical potential near the QH boundary is due to screening and interactions which flatten out the potential and density in the so-called wedding cake-like structure *(35,36)*. The width of the incompressible strips, *a*, can be estimated using Eq. S25 in Ref. *(36)*,

$$a = \left(\frac{4(4\pi\epsilon_0)\epsilon_r \Delta E_{\text{LL}}}{\pi^2 e^2 \frac{\partial n}{\partial y}}\right)^{\frac{1}{2}} \qquad S3$$

Here $\epsilon_0$ is the vacuum permittivity, $\epsilon_r$ is the relative permittivity, $e$ is the elementary charge, $\Delta E_{\mathrm{LL}}$ is the energy gap between LLs, and $\partial n/\partial y$ is the density gradient at the strip position. For the density gradient we can use the electric field profile across the boundary measured at zero field (Fig. S6). Here the boundary width is observed to be on the order of 70 nm, with a potential gradient of ≈1.7 meV/nm across the boundary (Fig. S6B). For the Dirac dispersion, $E = \hbar v_\mathrm{F} k_\mathrm{F}$, the density $n = E^2/\pi\hbar^2 v_\mathrm{F}^2$, and density gradient,

$$\frac{\partial n}{\partial y} = \frac{2E \frac{\partial E}{\partial y}}{\pi\hbar^2 v_\mathrm{F}^2}, \qquad S4$$

where $\hbar$ is Planck's constant divided by $2\pi$. Using $v_\mathrm{F} = 1.0 \times 10^6$ m/s, $\epsilon_r = 5$, the measured potential gradient $\frac{\partial E}{\partial y} = 1.7$ meV/nm and a value of $E = 40$ meV in the middle of the LL(0) to LL(-1) energy gap of $\Delta E_{\mathrm{LL}} = 80$ meV at $B = 5$ T, the density gradient from Eq. S4 is ≈ $10^{23}$ m$^{-3}$. Substituting the density gradient and 80 meV for energy gap in Eq. S3 yields an incompressible strip width of 34 nm, in good agreement with the measured strips in Fig. 4 of the main text and Fig. S5.

Moiré Superlattice

The rotational misalignment between the graphene and the hBN insulator underneath creates a moiré superlattice, which can have important consequences for the ground state of quantum Hall states as discussed in the main text. The moiré superlattice is observed in real and reciprocal lattice STM images of the graphene device as shown in Fig. S7, A and B. In Fig. S7A, both the atomic graphene lattice and the larger periodic moiré superlattice are observed simultaneously. The misalignment angle can be estimated by the moiré wavelength to be ≈3.1 degrees (*45*).

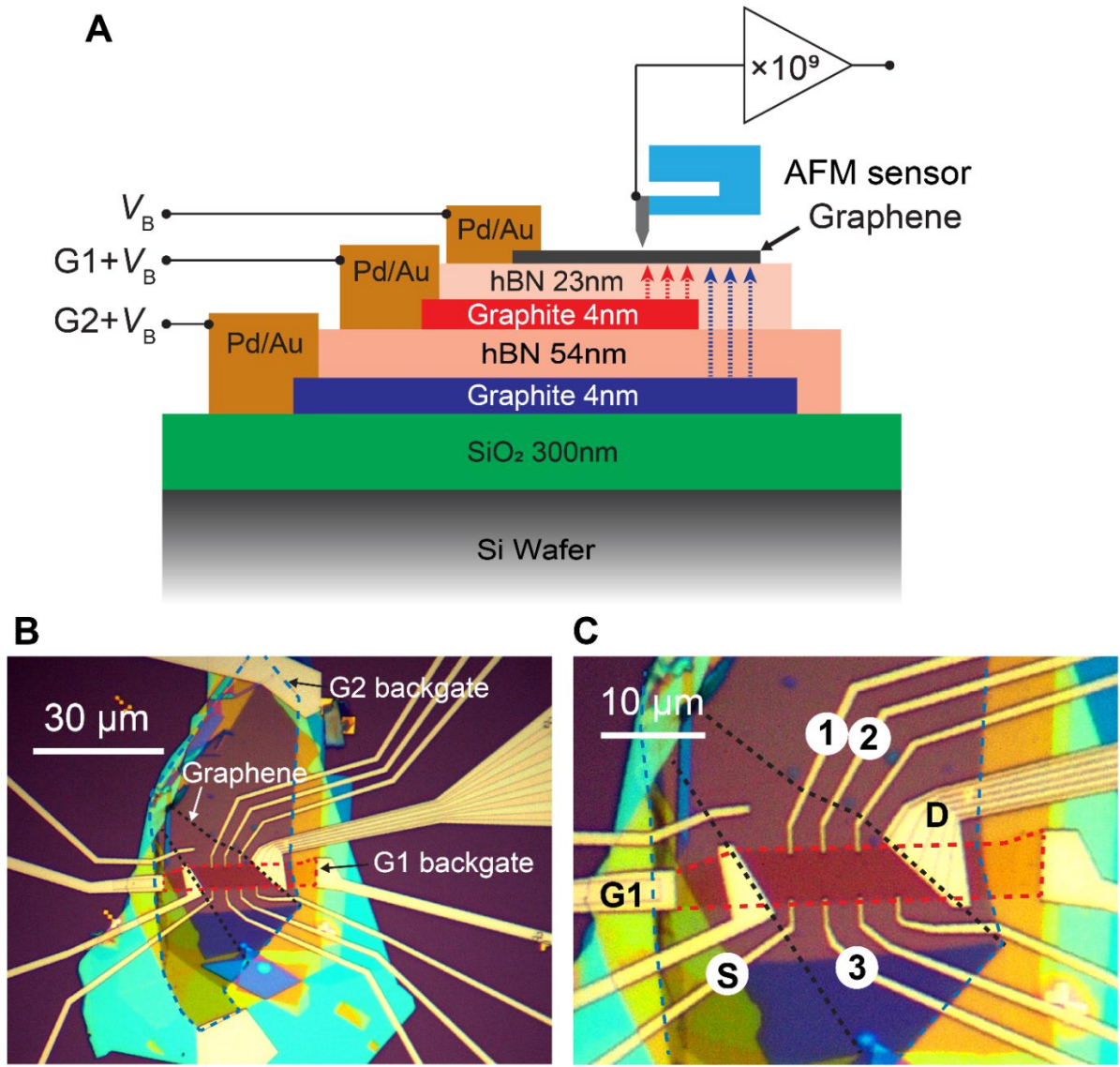

**Fig. S1.**

**Graphene quantum Hall device structure**. (**A**) Cross-sectional schematic of the layered structure of the graphene quantum Hall device. Two graphite back gates define the Hall bar: a global graphite back gate G2 (blue) and a local graphite back gate G1 (red). Pd/Au contacts are used to apply the sample bias $V_B$ to the graphene layer. (**B**) Optical image of the graphene device. The graphene sheet is indicated by the dashed black contour. The region controlled by the local gate G1 is shown by the dashed red line while the region controlled by global gate G2 is shown by the dashed blue line. Part of the fan "runway" used for approach is seen on the right side of the image. (**C**) Zoomed image of device region of (B) with the contacts indicated for transport measurements: source (S), drain (D), longitudinal resistance (1, 2), and Hall resistance (2, 3). The sample bias, $V_B$, is added to all contacts as shown in (A).

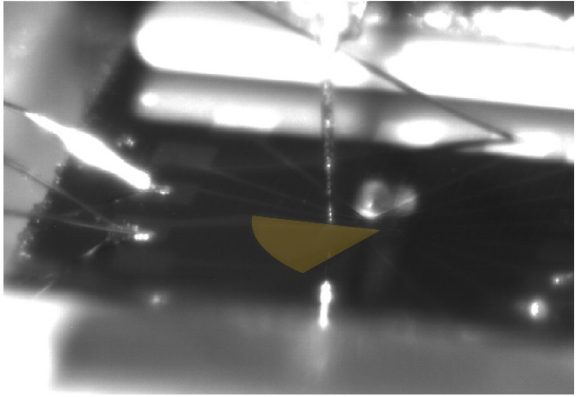
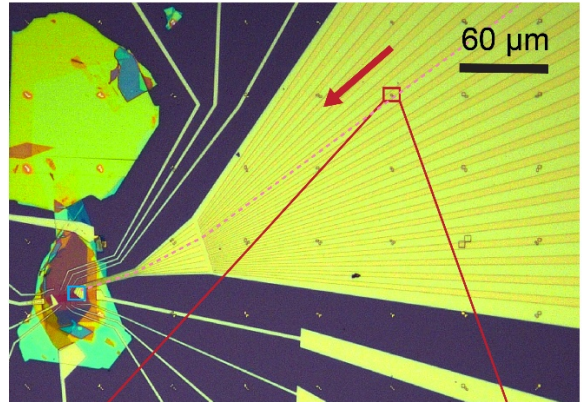
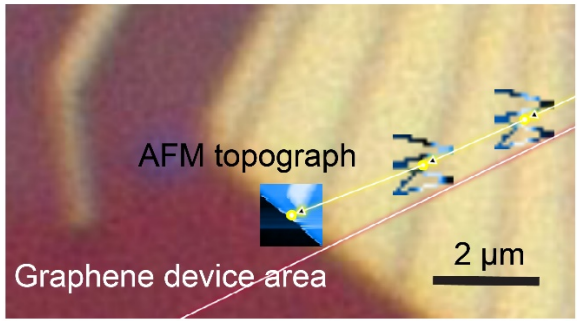
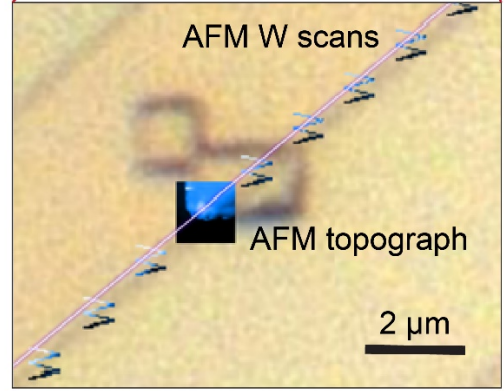

**Fig. S2.**
**SPM navigation to central device area**. (**A**) Optical image of the probe tip and its reflection as it is aligned onto the graphene device. The fan-shaped runway is artificially highlighted in yellow. (**B**) Optical image of the fan-shaped runway and device area (on the left). The red arrow indicates the navigation direction and pink dashed line shows the navigating pathway. (**C**) Zoomed-in optical image of a part of the navigation path. An AFM image and AFM "W"-shaped scans along the navigation path are superimposed. As shown, the automated algorithm repeats "W"-shaped AFM linescans to sense the ridge edge and re-calculate the next $X$ and $Y$ piezo-motor steps to adjust the direction. Along the navigation path, several full-frame scans are taken as shown to reveal markers to ensure accurate alignment of the navigation system to the optical image. (**D**) Optical image of the contact edge of the runway displaying the final navigation to the graphene device area. The AFM image showing the contact-graphene area is superimposed on the optical image. AFM topographs in (C) and (D) are measured at 2 nm oscillation amplitude and set frequency shift of −30 mHz.

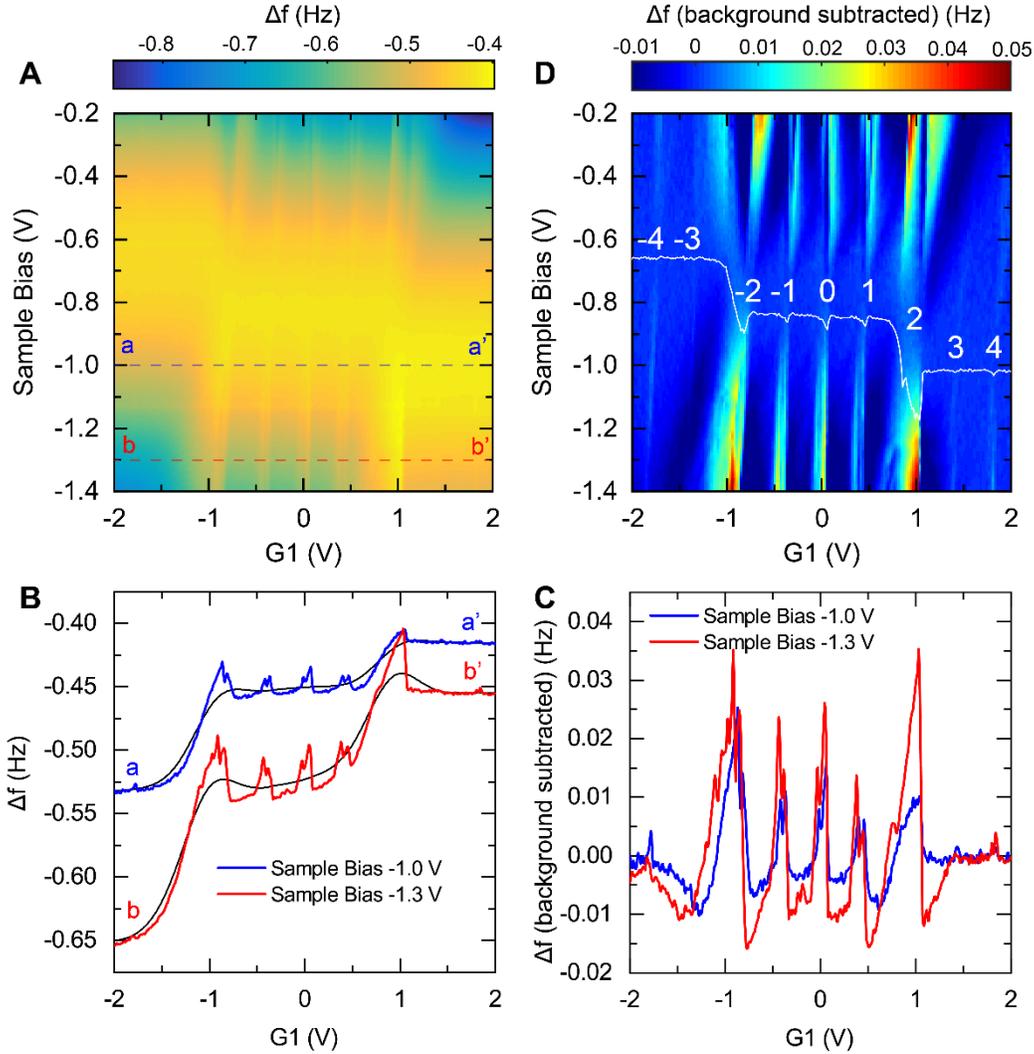

**Fig. S3.**

**AFM frequency shift measurement of broken-symmetry states.** (**A**) Raw frequency shift data as a function of sample bias and local back gate G1; the global back gate G2 is set to zero density. The frequency shift changes due to broken-symmetry states, seen as sharp spikes in the middle of the image, are superimposed on a pronounced background. The background is dominated by large steps corresponding to chemical potential changes at filling factors of $\nu = \pm 2$ along the gate axis, and electrostatic forces along the sample bias axis (see Fig. S4D). AFM data is measured at $B = 15$ T, with an oscillation amplitude of 2 nm at $\Delta f = -560$ mHz. The blue and red dashed lines indicate the positions for the linescans in (B). (**B**) Linescans from (A) along the gate axis at **a-a'** and **b-b'**, corresponding to sample biases of $V_B = -1.0$ V, and $-1.3$ V, respectively. The solid back line is a smooth background calculated for each linescan using a Gaussian filter with $\sigma = 0.2$ V. (**C**) Residuals obtained from subtracting the background curves shown in (B) from the raw data. (**D**) Frequency shift map obtained from the background-corrected linescans as in (C) for each sample bias, to highlight the changes due to the broken-symmetry states. The white line reflects the gate-dependent zero-contact potential, as obtained following the procedure in Fig. S4. The white numerals indicate the filling factor.

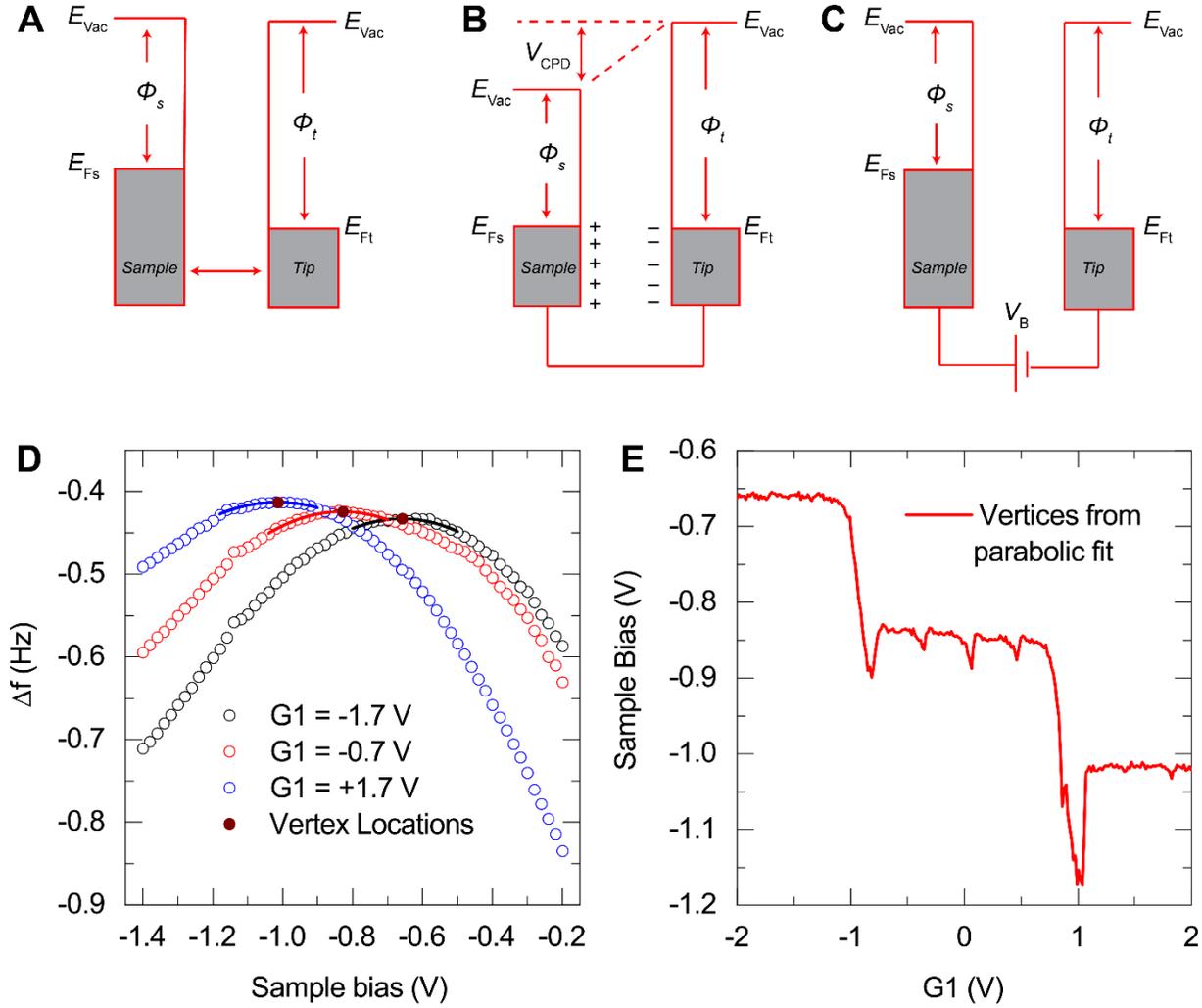

**Fig. S4.**
**Kelvin probe force spectroscopy of Landau levels.** Electronic energy levels of the probe tip and sample showing the Fermi level, $E_{Ft}$, $E_{Fs}$, vacuum level, $E_{Vac}$, and tip and sample workfunctions, $\varphi_t$, $\varphi_s$, for three cases: (**A**) Tip and sample are disconnected and separated by a large distance. (**B**) Tip and sample are in electrical contact resulting in charge flow and alignment of Fermi levels. A contact potential difference (CPD) is built up resulting in an electric field between the sample and tip. (**C**) A sample bias is applied between the sample and tip to null the CPD. (**D**) Linescans along the sample bias axis from Fig. S3A at selected gate voltages showing a downward parabolic response (symbols). The vertex of each parabola is obtained by fitting the data over a window of 0.3 V about the vertex (solid lines). The shift in the vertex positions is due to changes in chemical potential with gate voltage G1. (**E**) Vertex positions from fitting the downward parabolic response as in (D) for each gate voltage. The large steps are due to Landau levels at filling factors of $\nu = \pm 2$, while smaller steps are seen inside the $N = 0$ Landau level due to broken-symmetry states. Note that the chemical potential has a sign change along the sample bias axis as shown in Fig. 3 of the main text, where the KPFM data is obtained with lock-in measurements for improved signal to noise.

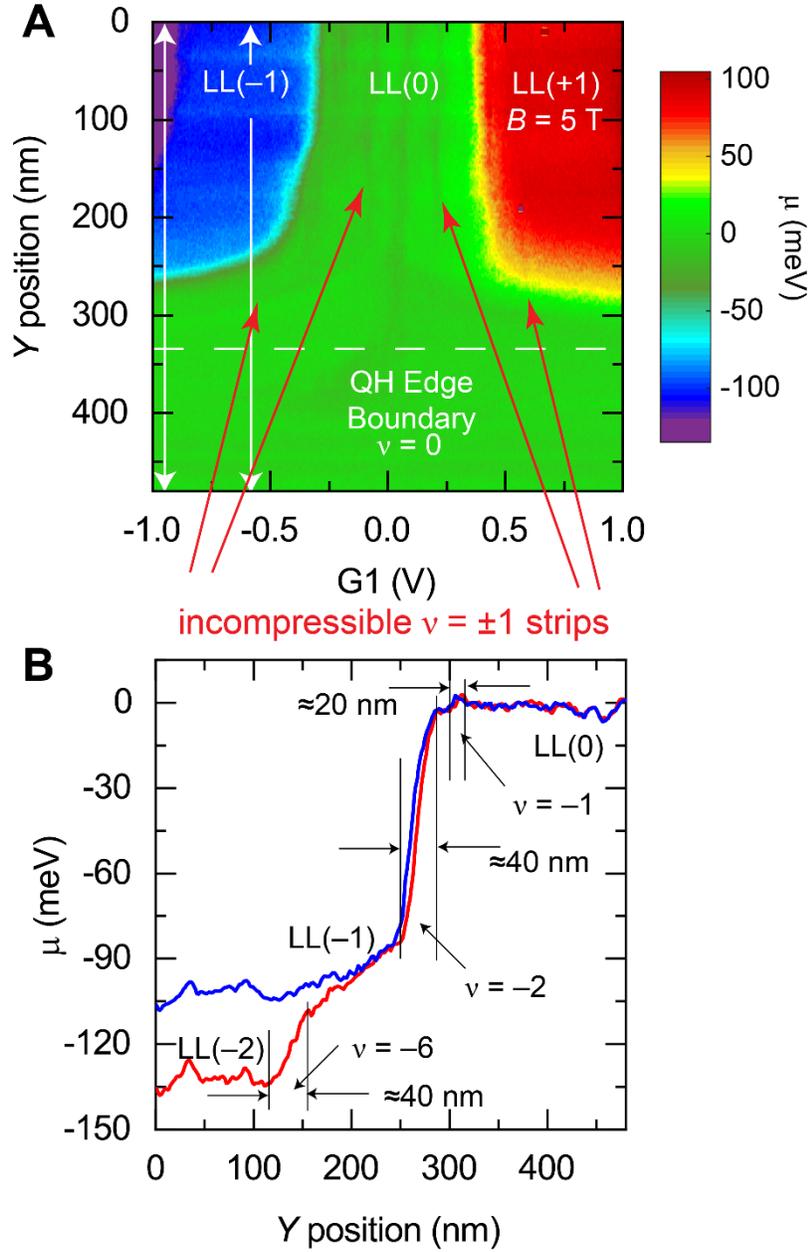

**Fig. S5.**

**KPFM measurements of incompressible strips.** (**A**) Kelvin probe map at $B = 5$ T of the chemical potential as a function of $y$-position across the quantum Hall boundary and local-gate potential, similar to the measurement at 10 T in Fig. 4 of the main text. In the local-gate area, Landau levels from $N = -2$ [LL(-2)] to $N = +1$ [LL(+1)] are seen in the different colored plateaus. AFM settings: 2nm oscillation amplitude, $\Delta f = -2$ Hz, and 20 mV sample bias modulation at 1.4 Hz. (**B**) Incompressible strips due to the large excursions in the chemical potential are observed in the lines traces at G1 = $-0.9$ V (red) and G1 = $-0.6$ V (blue) [white vertical lines in (A)] corresponding to filling factors $\nu = -6, -2$, and $-1$. The transitions separate plateaus between Landau levels, LL(-2) to LL(-1), and LL(-1) to LL(0).

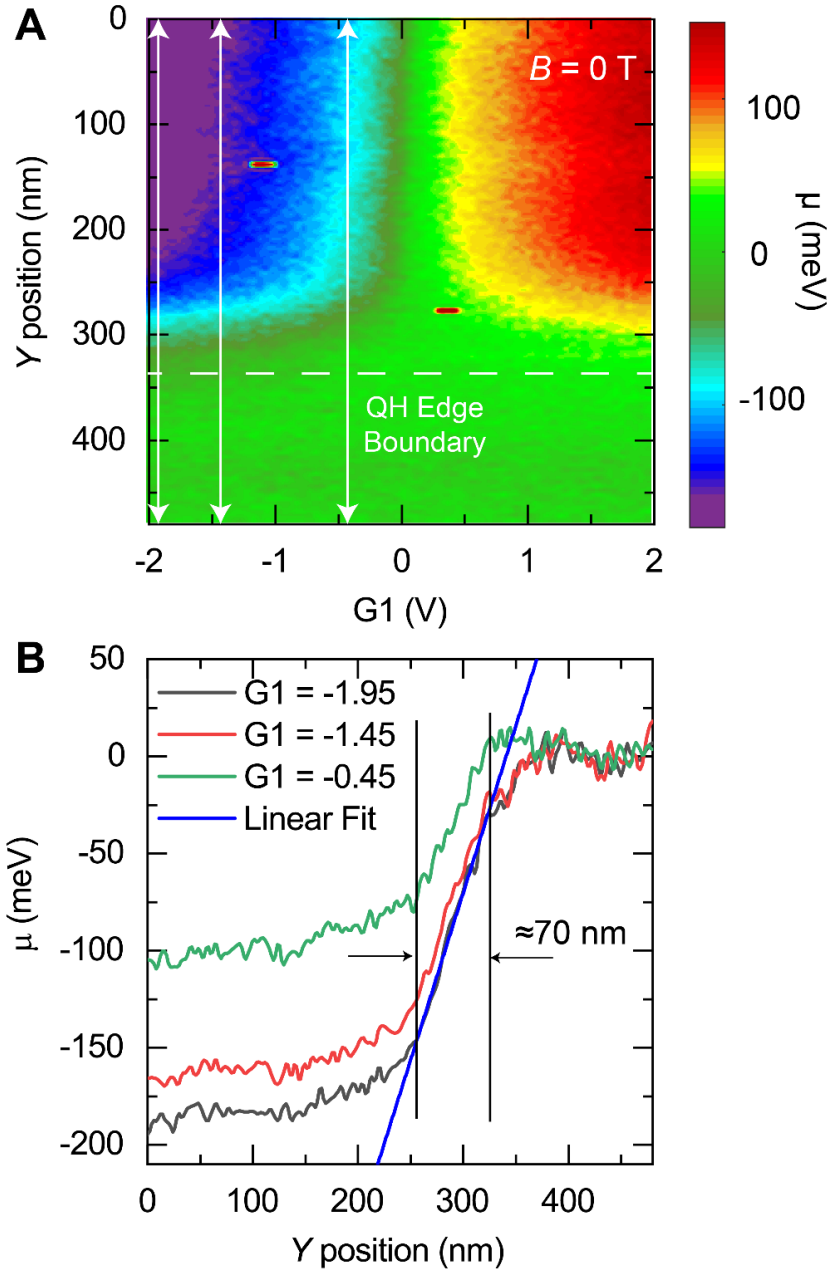

**Fig. S6.**
**Zero-field measurement of the boundary potential profile.** (**A**) Kelvin probe map at $B = 0$ T of the chemical potential as a function of $y$-distance across the quantum Hall boundary and local gate potential. The region outside the boundary was set to zero density with G2 = 100 mV. AFM settings: 2nm oscillation amplitude, $\Delta f = -2$ Hz, and 20 mV sample bias modulation at 1.5 Hz. (**B**) Line traces from (A) at the local gate potentials indicated by the solid white lines in (A), showing the width and sharpness of the potential boundary due to the use graphite back gates in close proximity to the graphene layer. A linear fit to the trace at G1 = −1.95 V (black) over the region bounded by the vertical lines yields a slope of $(1.72 \pm 0.02)$ meV/nm, where the uncertainty is one standard deviation from the linear least square fit.

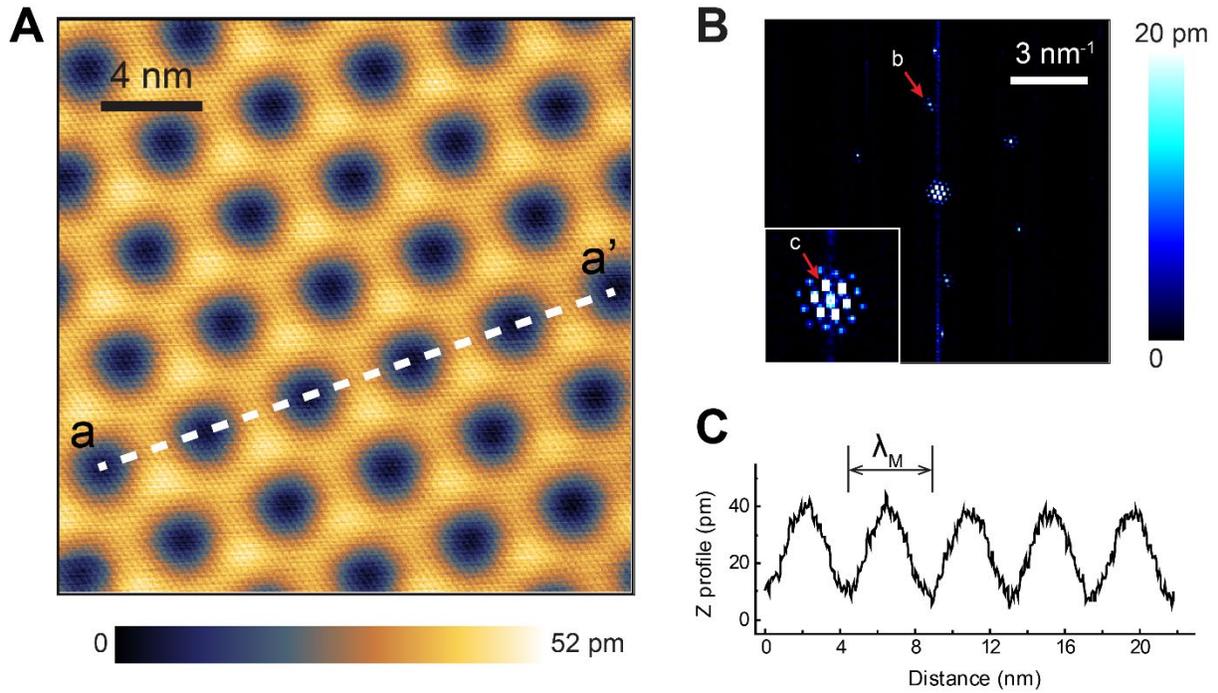

**Fig. S7.**

**Moiré superlattice of the graphene Hall device**. (**A**) STM topography of graphene surface. Dark and bright spots represent the moiré superlattice formed by the graphene sheet and the hBN underlayer. The graphene lattice is visible as the fine mesh in the whole area. Topography is obtained at $V_B = -100$ mV and a tunneling current of 300 pA. (**B**) FFT image of the topography image in (A). (inset) Zoomed view of the center area. Reciprocal lattice peaks of the graphene lattice are indicated by **b**. Reciprocal lattice peaks of moiré superlattice are indicated by **c**. (**C**) Z-height profile of the linecut **a-a'** in (A). The observed moiré wavelength $\lambda_M$ is 4.36 nm, which corresponds to a twist angle of ≈3.1 degrees between the graphene and hBN underlayer.